# Machine Learning-Assisted High-Throughput Semi-empirical Search of OFET Molecular Materials

Zhenyu Chen [a], Jiahao Li [c], Yuzhi Xu [b,d] *

a) School of Materials Science and Engineering, Nankai University, Tianjin 300350, P. R. China

b) School of Materials Science and Engineering, South China University of Technology, Guangzhou 510640, P. R. China

c) College of Chemistry, Nankai University, Tianjin 300071, P. R. China

d) Department of Chemistry, Brown University, Providence, RI 02912, USA

**Abstract:**
Machine learning has been widely verified and applied in chemoinformatics, and have achieved outstanding results in the prediction, modification, and optimization of luminescence, magnetism, and electrode materials. Here, we propose a deepth first search traversal (DFST) approach combined with lightGBM machine learning model to search the classic Organic field-effect transistor (OFET) functional molecules chemical space, which is simple but effective. Totally 2820588 molecules of different structure within two certain types of skeletons are generated successfully, which shows the searching efficiency of the DFST strategy. With the simplified molecular-input line-entry system (SMILES) utilized, the generation of alphanumeric strings that describe molecules directly tackle the inverse design problem, for the generation set has 100% chemical validity.  Light Gradient Boosting Machine (LightGBM) model's intrinsic Distributed and efficient features enables much faster training process and higher training efficiency, which means better model performance with less amount of data. 184 out of 2.8 million molecules are finally screened out with density functional theory (DFT) calculation carried out to verify the accuracy of the prediction. The influence of the screening criteria on the accuracy and efficiency of the prediction results is discussed. A reasonable error range in line with the current DFT calculation given as a standard, and this method is thousand times faster than high-throughput screening based on DFT. Here we propose the concept to actively generating high density search (HDS) of valid chemical space towards certain type of functional molecules, in which DFST is one of search strategies, machine learning methods as a filter, and combining DFT calculation for verification.

## I. Introduction

Organic field-effect transistors (OFETs) with many excellent qualities such as low cost, flexibility, bio-compatibility, has gain particular attention in the new technological evolution of electronic transistors[1]. Compared with traditional amorphous-silicon-based thin-film transistors (TFTs)[2-3], OFETs have irreplaceable advantages in active matrix/organic light emitting diode (AMOLED) circuit[4-5], passive RFID tags[6-7], and other relating fields[8]. Based on transport situation,

OFET materials as the key in OFETs can mainly subdivide into p-type, n-type, and am-bipolar semiconductors, respectively[9]. P-type OFETs have achieve much success for their peculiar geometry, high electron donating and hole transporting ability[10-11]. while n-type OFETs commonly adopted dimide-, amide-, and acene-based materials, have only achieved 5 $cm^2V^{-1}s^{-1}$ of electron mobility on average[12-13]. Compared to that of p-type, n-type OFETs are three orders of magnitude less. For example, P(NDI2OD-T2), a donor−acceptor (D−A) copolymer of 2, 6-naphthalenediimide and 2, 2'-bithiophene, by Li et al. [14] and Facchetti et al[15]. independently reveals many distinctive advantages, such as solution processing, relatively high electron mobilities up to 3.50 $cm^2 V^{−1} s^{−1}$. Further, stability of OFET in the air is also a problem remaining to resolve[16]. To realize the air-stable OFET, the organic semiconductors need to have HOMO energy levels which are lower than approximately -5.3 $eV^{17}$, which can not be reached by conventional organic semiconductors, such as pentacene (-5.0 eV)[18], poly(3-hexylthiophene) (P3HT, -4.7 eV)[19], and copper(II) phthalocyanine (CuPC, -5.0 eV)[20]. Therefore, n-type organic semiconductors can not meet the requirements of logic circuits and affect the stability of device performance. Among many factors, HOMO, LUMO, and bandgap are regarded as the most crucial physical parameters for excellent performance in stability and electron mobility of n-type OFETs.[21-22]

The HOMO energy level decided the air stability of the OFET materials and the proper gap between the HOMO and LUMO help avoid unexpected chemical reactions.. In previous work, DFT calculation plays an important role in the prediction and virtual screening to achieve a suitable HOMO, LUMO, Bandgap[23-25]. However, DFT cannot accurately describe the electronic structure of the excited state of the material, such as the low electronic band gap under LDA[26]. It's difficult to describe long-range weak interaction like van der Waals force, and for large systems and long-term scales, the amount of calculation is still a lot ($\sim N^3$)[27].

Machine learning approaches have been widely verified and applied in materials and chemistry informatics, and have achieved outstanding cases in the prediction and modification of luminescence[28-29], magnetism[30-31], and electrode materials[32-33], which have been verified by experiments[34-35]. Based on a large amount of data and machine learning algorithms, the structure-activity relationship is quantitatively established, which is helpful to guide experimenters in future synthesis, analysis and testing[36-37]. Based on the previous research, the HOMO and LUMO energy together with the $E_F$ (*Fermi energy*) of the source electrodes can be extracted or calculated from the accessible database. These main electronic properties of n-type materials are recently been proved by Min-Hsuan Lee et al. to have strong relationship with the experimental electron mobility of n-type OFETs, which, to be more specifically, gives the standard of molecules with a combination of the HOMO level of −5. 8 eV and the LUMO level of −3.75 $eV^{38}$. These researches provide the possibility to predict the electron mobility of n-type OFETs by using only desired electronic properties of n-type materials. To obtain a suitable novel OFET materials are difficult because the device and synthesis process should be in consideration in industry. The generator with GAN and other methods to generate the molecules are commonly applied in previous research of drug discovery[39-40]. However, these methods generate novel molecules in chemical space, which shows largely unique from known structures, It brings more challenges to apply these molecules to specific devices. Moreover, since the development from OFET materials to devices application suffer from a long period, how to combine precise first-principle calculation with high-precision screening of molecules to accelerate the discovery of OFET still needs to be explored.

With training standard from the machine learning model together with other empirical rules from previous researches, HOMO and LUMO energies can be optimized perfectly to improve the OFETs'electron injection efficiency together with the ambient stability against atmospheric oxygen and moisture. This work established the research paradigm of "Traverse the molecular structure of the specified skeleton—machine learning screening—density functional theory and reorganization energy screening—target high-performance molecules" and completed the screening of N-doped and functionalized molecules of tetracene and pentacene, and verified the relationship between the LUMO and HOMO energies and electron transport performance. Compared with pure density functional theory calculation and screening, the machine learning together with density functional theory material exploration mode has the following advantages: (1) Depth first search traversal algorithm and molecular fingerprinting can maximize possible molecular structures and enrich the structure of the candidate material (2) The machine learning algorithm model can quickly and effectively predict and initially screen the molecular structure of the structural skeleton type material with high electron transport performance through the HOMO and LUMO energies and empirical rules (3) The secondary fine screening of DFT calculation and reorganization energies calculation enable the model to improve the accuracy of data prediction while ensuring the screening speed, and feedback to the machine learning model for multiple improvements, and finally achieve high-efficiency and high-precision prediction.

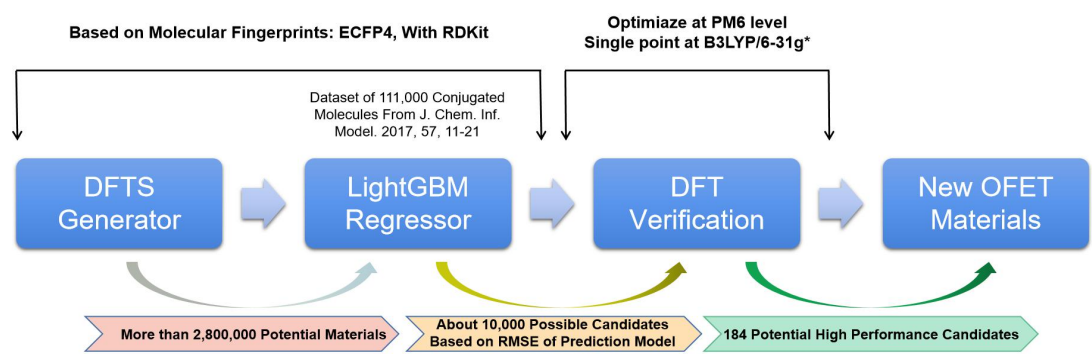

**Figure 1.** The process of OFET molecular material discovery. Totally three steps of generation, regression, and verification before we discover new OFET materials.

In this work, we propose the concept to actively generating high density search (HDS) of valid chemical space towards certain type of OFET materials, which in this case DFST is one of search strategies, machine learning methods as a filter, and combining DFT calculation for verification (Figure 1). Different searching approaches will influence the searching results, and here we utilize the DFST approach combined with lightGBM machine learning model as an example to search the classic Organic field-effect transistor (OFET) functional molecules chemical space, which is simple but effective. Totally 2,820,588 molecules of different structure within two certain types of skeletons are generated successfully, which shows the searching efficiency of the DFST strategy. We use the simplified molecular-input line-entry system (SMILES) to generate alphanumeric strings that describe molecules. This method tackles the inverse design problem, for the generation set has 100% chemical validity. Light Gradient Boosting Machine (LightGBM) model's intrinsic Distributed and efficient features enables much faster training process and higher training efficiency, which means better model performance with less amount of data. 184 out of 2.8 million molecules are finally screened out with density functional theory (DFT) calculation carried out to verify the accuracy of the prediction. Then we discuss the influence of the screening criteria on the accuracy and efficiency of the prediction results. A reasonable error range in line with the current DFT calculation given as a standard, and based on the data set of the same magnitude, this method is thousand times faster than high-throughput screening based on DFT.

## II. Methods

**Deep First Search Traversal Generator.**

The depth first search traversal of the graph is to start from a vertex v in the graph: (1) Visit the vertex v; (2) Start from the unvisited neighboring points of v in turn, and perform depth-first traversal on the graph until the vertices with the same path are visited; (3) If there are still vertices in the graph that have not been visited at this time, start from an unvisited vertex, and re-execute the depth-first traversal until all vertices in the graph have been visited. Here, we regard the arrangement of atoms in the molecule as a graph structure, and perform depth-first traversal.

During training, the atom was represented with ECFP4 of RDKit. After converting them back to SMILES, every constituting character was one-hot encoded. Every SMILES string was thus represented by a two-dimensional (2D) array. Next, two-step deep first search traversals were performed based on tetracene and pentacene. In the first traversal, the carbon atoms in the skeleton are replaced with nitrogen atoms, and the upper limit of replacement is 3. After removing the same structure representation, a total of 452 skeleton structures were generated. Based on this, we performed a second traversal. five types of functional groups were traversed at most 2 substitution sites according to 3 connection ways, and finally 2820588 molecules were generated.

**LightGBM Model.**

LigthGBM is a new member of the boosting ensemble model provided by Microsoft. It is an efficient implementation of GBDT like XGBoost. In principle, it is similar to GBDT and XGBoost. Both use the negative gradient of the loss function as the residual approximation of the current decision tree. To fit a new decision tree. LightGBM will perform better than XGBoost in many aspects. It has the following advantages: faster training speed and higher efficiency, lower memory usage, better accuracy, support of parallel, distributed, and GPU learning, and capable of handling large-scale data. In a large sample and high-dimensional environment, traditional boosting cannot meet current needs in terms of efficiency and scalability. The main reason is that traditional boosting algorithms need to scan all sample points for each feature to select the best, which is very time-consuming. In order to solve this time-consuming problem in the environment of large samples and high-latitude data, LightGBM uses the following two solutions: One is GOSS (Gradient-based One-Side Sampling). Instead of using all the sample points to calculate the gradient, GOSS samples some of the sample to calculate the gradient. The second is EFB (Exclusive Feature Bundling), here is not to scan all the features to obtain the best segmentation point, but bundling certain features together to reduce the dimensionality of the features and find the best segmentation point to reduce the consumption. This greatly reduces the time complexity of processing samples, but a large number of experiments have proved that using LightGBM on some data sets does not lose accuracy, and sometimes even improves accuracy.

**DFT Calculation.**

About 10 thousand molecules are optimized at PM6 level and calculate the single point at B3LYP/6-31g*.

**Datasets.**

The number of molecules produced by the generator and the control hierarchical filter is shown in Table 1. Through depth-first traversal and preset skeleton structure composition, 2820588 molecules are generated. The machine learning model screening uses lightGBM model to predict the HOMO and LUMO energy of the molecules. The data set after the initial screening will be verified by the DFT secondary screening after structure optimization. We then analyzed and discussed the selection of the data set after the machine learning model prediction, and determined the optimal parameter $\gamma = 1.98$. The selected molecules are shown in the following figure xxx

## III. Results and Discussion
### Depth First Search Traversal (DFST) Generator

In the discovery process of existing drugs or other molecular materials, researchers often follow existing empirical rules to artificially optimize the classical structure, or adopt different features to machine learning for structural optimization, or use existing molecular database to make property predictions. These methods have their domain rationality based on the quantitative structure-property relationship (QSPR).. In many cases of drug discovery, adopting a generator model to create new structures are desired with clear target functional group. However, in the realm of electronic materials applied in devices , many factors should be taken into consideration, such as devices structures and the complexity and feasibility of material synthesis. For example, in organic solar cells (ORCs) discovery, to maximize the cell efficiency, the whole device processing needs to be re-examined and optimized so as to apply the new materials. As a result, thorough study on organic materials' physical background is necessary. It is hard to ensure the new kind of material discovered from a relatively large chemical space via a normal generation methods. Discovery in OFET materials have empirical knowledge from previous works, from which we can benefit the discovery process and modify the generation method . Using the familiar molecular skeletons greatly reduces the workload of adjusting the device preparation process, and is conducive to the practical application of OFET materials. Moreover, methods of GAN or Reinforcement Learning Methods conduct training and data generation based on    former molecule structures dataset, which means it is difficult to conduct unbiased searches in the certain range of the chemical space due to empirical biases or unknown errors, resulting in inefficiency and omissions in material discovery. In addition, due to the relatively existing molecular data sets Since the entire chemical space is still in a small-scale data set, direct structural optimization on this basis can easily lead to larger errors and biases from the data set. So, Here, we proposed our new generator in OFET, which is named as Depth First Search Traversal (DFST)

Generator.

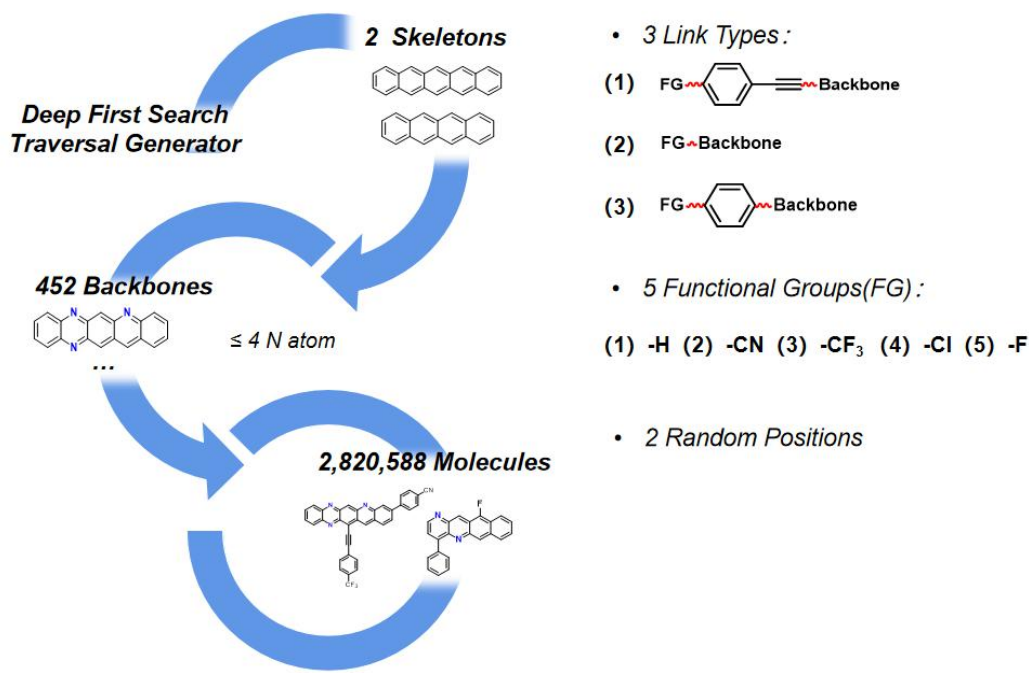

**Figure 2.** The details information of Deep First Search Traversal (DFST) Generator, including the 3 link types, 5 funtional groups and 2 random positions.

For functional molecules with a relatively fixed skeleton in a specific field, we propose to use a traversal method to search and traverse the chemical space with high density and efficiency. Here we use the deep first search traversal based on 2 basic skeletons to carry out the atoms of the skeleton. Naphthacene and Pentacene are chosen here as the basic skeletons, one fundamental reason is that acenes with rich aromatic cores, are born to be the excellent hole transporting materials. The aromatic system has a strong impact on the LUMO energy of the entire system. Another important reason is that some benchmark researches have been carried out before. Moreover, the nitrogen atom are chosen to form N-heteroacenes because it is regarded as a highly promising n-type acene skeleton semiconductor with abundant researches before. The synthesis and transistor performance of them are guaranteed from the reference[1]. And the researches have proved the influences of different nitrogen positions in the backbones of the molecules on the molecule packing and electronic behavior. With that says, changing nitrogen atom substitution positions and number will greatly tune the material behavior, which is an efficient searching action. The typical functional group -H, -CN, -CF3, -Cl, -F are chosen because of their electron withdrawing quality. With the electron deficiency induced, the chemical space of n-type materials with electric charge density adjusted are fully traversed. As a result, about 2.8 million molecular structure datasets are generated. Then we chose 3 link types for these 5 functional groups (Figure 2) to substitute the skeletons in two random positions. The depth-first search traversal method is based on the experimental empirical knowledge and can generate a variety of skeleton structures, and is suitable for molecular materials generation tasks with multiple constraint conditions in various fields.

**LightGBM Regressor**

As DFST generator provides a large magnitude of data set, traditional high-throughput screening is unacceptable due to the high cost and long time. Thus, several machine learning algorithms as an initial screening method, and lightGBM model and ECFP4 with RDKit was finally chosen as an suitable screening method to create a fast and efficient initial screening (Figure 3a). Machine learning method are based on molecular descriptors and some other supplementary descriptors, which can give predictions based on the potential relationship model of molecular structure and LUMO and HOMO data in existing databases.

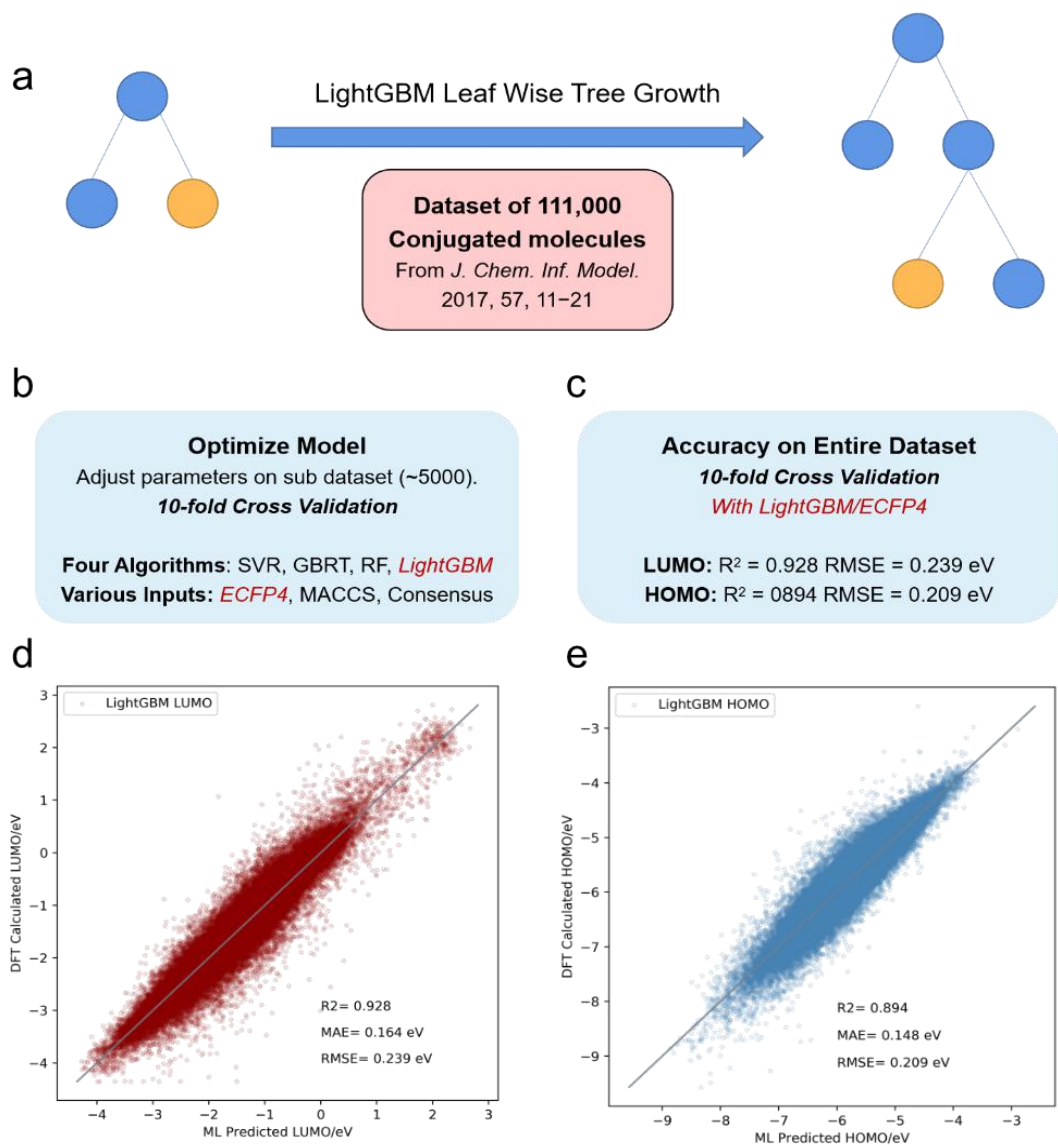

**Figure 3**. (a) The schematic diagram of LightGBM. (b) Model optimization process. (c) Prediction results of optimal model (LightGBM + ECFP4). (d) The correlationship of DFT calculated LUMO and ML predicted LUMO. (e) The correlationship of DFT calculated HOMO and ML predicted HOMO.

As is shown in the Figure 3b, based on a dataset of 111,000 conjugated molecules, four algorithms and various inputs were screened to adjust parameters on subdataset(~5000) with a 10-fold cross validation. Based on the molecular fingerprints and ECFP4 with RDKit, the regressor utilize GBRT to predict electronic properties. Then we use lightGBM&ECFP4 to maintain the accuracy on entire dataset and finally get root mean square error (RMSE) up to 0.21 and 0.24 eV for the HOMO and LUMO orbitals and $R^2$ up to 0.894 and 0.928, respectively (Figure 3c to 3e). Moreover, because of a leaf-wise tree growth strategy and novel techniques, LightGBM has been proved to to be up to 20 times faster using the same training set, compared to the XGBoost implementation of GB. As a result, based on the HDS generator and ultra-fast regressor, more than 2,800,000 potential materials are screened into about 10,000 possible candidates with a standard of LUMO and HOMO energies from the formal research ( 3.75±0.3eV and 5.80±0.3eV, respectively ).

**DFT Secondary Screening**

The number of potential target molecular groups obtained through the preliminary screening of machine learning methods is still large, and there is still a large accuracy gap compared with the solution theory calculated by DFT. For this reason, we performed DFT on the 10,000 small molecules selected. The HOMO and LUMO energy and the reorganization energy were calculated, so as to perform secondary screening on the preliminary screening data set according to the standard. The importance of secondary screening is multiple: (1) The screening results can be used to strengthen the training of the original model, thereby improving the accuracy of the original model, thereby continuously improving the model and completing reinforcement learning (2) It can verify the scientific nature of machine learning algorithm screening (3) It can further reduce the number of target molecules and improve the prediction accuracy, which can specifically and effectively guide experimental synthesis and analysis. After optimized at PM6 level and calculate the single point at B3LYP/6-31g*, we finally find 184 target molecules.

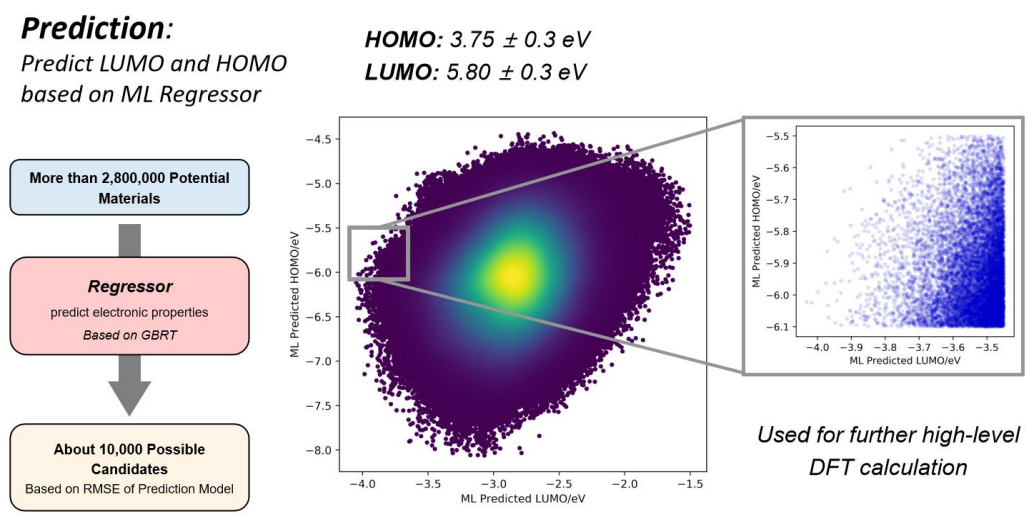

**Figure 4.** The prediction part and principles for selecting data. HOMO between 3.75±0.3eV and LUMO between 5.80±0.3eV are selected for further high-level DFT calculation.

The representative molecular structure is shown in Figure 5 below, and the HOMO, LUMO energy based on the LightGBM model prediction and DFT verification results are listed in the table. It can be seen that DFST Generator successfully generated different structures that meet the energy requirements of HOMO and LUMO based on tetracene and pentacene, and has a certain similarity with the structure of the previous traditional OFET molecular materials, which is synthesizable and maintain stability in device in the subsequent work. It provides kind of guarantee for the electronic device material discovery.

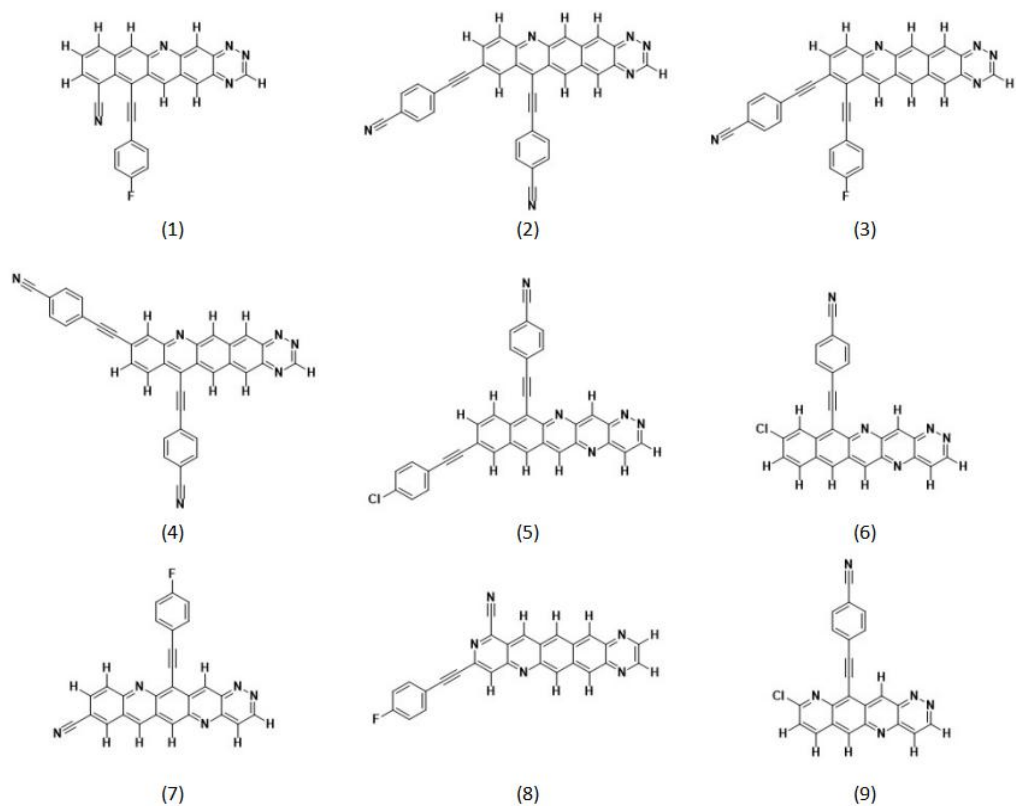

**Figure 5.** Selected molecular structures of DFST generated OFET molecular materials.

**Table 1.** Prediction on HOMO and LUMO energy based on the LightGBM model and DFT verification results.

| Mol | LightGBM LUMO(eV) | LightGBM HOMO(eV) | DFT LUMO(eV) | DFT HOMO(eV) |
| --- | --- | --- | --- | --- |
| 1 | -3.76 | -5.82 | -3.69 | -5.53 |
| 2 | -3.77 | -5.79 | -3.67 | -5.67 |
| 3 | -3.73 | -5.78 | -3.55 | -5.52 |
| 4 | -3.72 | -5.80 | -3.66 | -5.67 |
| 5 | -3.78 | -5.79 | -3.70 | -5.58 |
| 6 | -3.75 | -5.81 | -3.74 | -5.64 |
| 7 | -3.75 | -5.79 | -3.72 | -5.58 |
| 8 | -3.73 | -5.80 | -3.56 | -5.60 |
| 9 | -3.74 | -5.81 | -3.55 | -5.94 |

**Discussion on the Standard of Precise and Efficient Screening**

For molecules that meet the conditions of LUMO and HOMO after being screened by the machine learning model, DFT calculations will be called to perform more accurate and reliable traditional calculations on the frontier orbits of the molecules. In the previous researches, the computational scientists have focused on various factors that may have impacts on the performance of the machine learning model. While in the filed of materials chemistry, due to the relatively small scale of the training set, the former discussion mainly concentrates on the size and the contents of the training set. While different size and contents of training set may cause either positive or negative effects on the performance of the imbalanced learning model, there is few researches on the criteria of optimizing the parameters in the materials prediction or classification models. However, this is of much significance for many reasons. Considered that the virtual screening results have to be verified by classic time-costly quantum chemical methods like DFT, and also the traditional DFT methods still maintain acceptable error which is called chemical accuracy, the parameters in the model should not only promise the proper accuracy of the target prediction missions, but also filter out enough data to improve the whole discovery efficiency and the computational verification cost. Root mean square error (RMSE) is a commonly used measure of the difference between the predicted values. Here we propose to screen the generation set with the predicted data of the lightGBM model together with the RMSE of the prediction results. RMSE reflects the prediction inaccuracy scale. Combined with the thresholds from the chemical knowledge, it can be used as a simple tool to rationally control the size of the screened-out molecules. Due to the inherent accuracy limitations of the DFT calculation, adjusting the acceptable error range of the machine learning model prediction result too small will cause its ideal accuracy to deviate from the actual accuracy requirements, and make the number of molecules obtained by screening too small, but setting the error result predicted by the machine learning model to be too large will result in a screening efficiency that is too low.

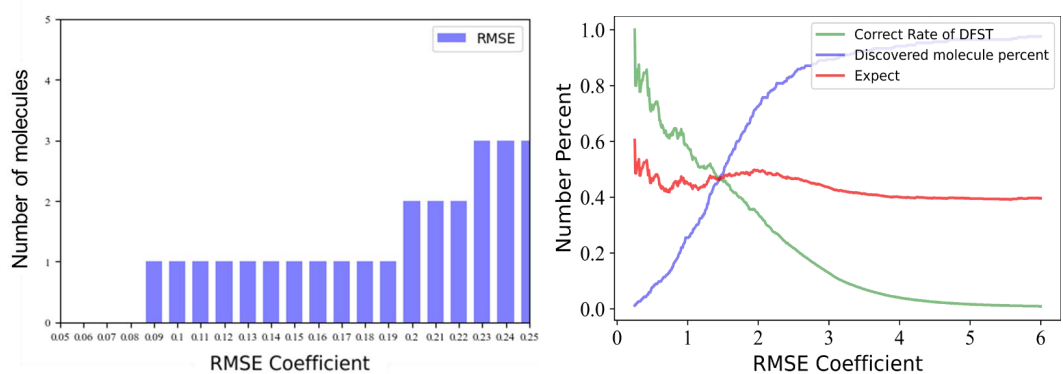

**Figure 6.** Discussion on the balance between the discovery ability and efficiency. **(a)** Number of molecules when the RMSE Coefficient are controlled within 0.25. **(b)** The Expect of the RMSE coeffcient to achieve a relatively high correct rate of DFST while maintaining high discovery efficiency.

As can be seen from the Figure 6a when we control the data error of the machine learning model within the range of 0.25RMSE, only less than 10 molecules can meet the condition and be retained by us. In fact, too few molecules actually indicate that potentially satisfying molecules are eliminated by unnecessary precision constraints. At the same time, the figure 6b reflects that when we relax the accuracy requirements of the machine learning model, the number of molecules that need DFT verification will increase rapidly. This will be followed by a substantial increase in time and computing power costs, as well as large errors. The rapid decline in the screening accuracy rate leads to a waste of costs caused by unqualified small molecules. Considering the model error compensation and the acceptable accuracy of DFT theoretical calculation, we will find the efficiency and the arithmetic average of the correct rate of one screening as the objective function. When the expectation reaches the maximum, we believe that the model efficiency is the highest at this time. Therefore, we get the reasonable RMSE coefficient = 1.98, which promise a comparatively high screening efficiency and also high data accuracy.

## IV. Conclusion

Here we propose a deep first search traversal (DFST) molecule generator based on the empirical knowledge to explore tunable and synthesizable n-type OFET materials. Combined with lightGBM machine learning model, 2880588 molecules of different structures input helps quickly pick up about 10,000 molecules by precise prediction of the HOMO and LUMO energy. 184 out of 1 million molecules are finally screened out with density functional theory (DFT) calculation carried out to verify the accuracy of the prediction. To get a balance between the discovery ability of lightGBM and the discovery efficiency, we design an expect function and discuss the rationality of the formula. A reasonable RMSE coefficient in line with the current DFT calculation was given as a standard, and this method is thousand times faster than high-throughput screening based on DFT. It is worth mentioning that the DFST strategy can effectively search the chemical space with limited empirical knowledge, which achieves a balance between empirical bias on traditional machine learning training access and the correct instruction of experimental experiences. And the utilization of the lightGBM model can filter out the target molecule up to 3‰ of the original data set within 1/1000 of the time of DFT calculation, and subsequent verification can ensure the scientific insight and accuracy of the data. We believe that similar materials generation and discovery paradigms are not only suitable for the discovery of OFET materials, but also have a general application prospect on the design and discovery of other molecular materials and various one-dimensional and two-dimensional materials.

## AUTHOR INFORMATION

**Corresponding Author**

* E-mail: yuzhi_xu@brown.edu (Y.Xu)

**Notes**

The authors declare no competing financial interest.

**Acknowledgements**

We gratefully acknowledge the financial support of C.W. Server Plan. We thank Prof. Lianjie Zhang at SCUT for helpful discussion.